\documentclass[11pt,twoside]{article}
\usepackage{CAGN2019}
\usepackage{graphicx}
\usepackage[T1]{fontenc} % Computer Modern (CM) fonts
\usepackage{latexsym}
\usepackage{verbatim}
\usepackage{ifpdf}
\usepackage{amsmath,amssymb}
\ifpdf
      \DeclareGraphicsExtensions{.pdf,.png,.jpg}
\else
      \DeclareGraphicsExtensions{.eps}
\fi
\newcommand\ion[2]{#1$\;${\scriptsize\rmfamily{#2}}\relax}

\begin{document}
% please do not un-comment the next line
% \input{../../proceeding-book/expages.tex}\setpagenumber{1}

\vskip 1.0cm
\markboth{Manuel Peimbert}{50 YEARS OF TEMPERATURE INHOMOGENEITIES IN GASEOUS NEBULAE}
\pagestyle{myheadings}
%
%%%%  USE THE LINE THAT DESCRIBES THE CHARACTER OF YOUR WORK %%%%%%
%
\vspace*{0.5cm}
\parindent 0pt{Invited Review}
%\parindent 0pt{Contributed  Paper}
%\parindent 0pt{Poster}
%\vskip 0.3cm

\vspace*{0.5cm}
\title{50 YEARS OF TEMPERATURE INHOMOGENEITIES IN GASEOUS NEBULAE}

\author{Manuel Peimbert}
\affil{Instituto de Astronom\'ia, Universidad Nacional Aut\'onoma de M\'exico, Apdo. Postal 70-264 Ciudad Universitaria, M\'exico}

\begin{abstract}
A review of some of the papers that discuss the relevance of the temperature structure present in \ion{H}{II} regions and planetary nebulae is presented. Particular attention is given to the determination of the chemical abundances of these objects.
\bigskip


\end{abstract}

\section{Introduction}

Some of the following ideas are the result of my collaboration with: Silvia Torres-Peimbert, Antonio Peimbert, Valentina Luridiana, Rafael Costero, Julie\-ta Fierro, Miriam Pe\~na, Leticia Carigi, Gloria Delgado-Inglada, C\'esar Esteban, Jorge Garc\'ia-Rojas, Mar\'ia Teresa Ruiz, Charles Robert O'Dell, and Gary Ferland.

The determination of the chemical abundances of \ion{H}{II} regions and Planetary Nebulae is paramount for the study of the evolution of the stars, the galaxies and the universe as a whole. The determination of the  chemical abundances strongly depends on the temperature structure
of the gaseous nebulae.

This review is based on two papers the first one \citep{Peimbert1967} discusses the differences in the electron  temperatures derived from different methods, and the second one \citep{Peimbert1969} applies the formulation of the first paper to derive the chemical composition of gaseous nebulae.

A recent review on the determination of the physical conditions and chemical abundances of gaseous nebulae has been presented by \citet{Peimbert2017}.

\section{Temperature determinations}

The average temperature of a gaseous nebula is given by

\begin{equation}
T_{0}(X^{i+})=\frac{\int T_{e}n_{e}n(X^{i+})dV}{\int n_{e}n(X^{i+})dV}.
\end{equation}

The mean square temperature fluctuation is given by

\begin{equation}
t^{2}(X^{i+})=\frac{\int \left[T_{e}-T_{0}(X^{i+})\right]^{2}n_{e}n(X^{i+})dV}{T_{0}(X^{i+})^{2}\int n_{e}n(X^{i+})dV}.
\end{equation}

The temperature derived from the [\ion{O}{III}] lines is given by

\begin{equation}
T_{4363/5007} = T_{0}\left[1+\frac{t^{2}}{2}\left(\frac{91300}{T_{0}}-2.68\right)\right].
\end{equation}

The temperature given from the ratio of the Balmer continuum to H beta is given by

\begin{equation}
T_{Bac} = T_0\left(1 - 1.67 t^2\right).
\end{equation}

\section{Presence of temperature Inhomogeneities}

For the best observed objects it is found that the forbidden line temperatures are higher than the Balmer temperatures, the differences are higher than  those predicted by the   theoretical photoionization models. From photoionization models the $t^2$ values predicted by \texttt{CLOUDY} for PNe and \ion{H}{II} regions are in the $0.000$ to $0.015$ range with a typical value of about $0.004$ \citep{Ferland2013, Ferland2017}.

\citet{Peimbert2012} from 37 galactic and extragalactic \ion{H}{II} regions found $0.019<t^2<0.120$ with an average value of $0.044$.

\citet{Toribio2017} from 5 \ion{H}{II} regions in the LMC found $0.028<t^2<0.069$ with an average value of $0.038$, and from 4 \ion{H}{II} regions in the SMC found that $0.075<t^2<0.107$ with an average value of $0.089$.

The abundances relative to hydrogen derived from recombination lines of oxygen and carbon are almost unaffected by temperature variations.

The abundances derived from collisionally excited lines, under the assumption of constant temperature, typically underestimate the abundances relative to hydrogen by a factor of 2 to 3 (C, O, Ne), the so called abundance discrepancy factor, ADF. By adopting $t^2$ values different from $0.00$ it is possible to reconcile the abundances derived from forbidden lines
with those derived from permitted lines.

\section{Planetary Nebulae}

There are two families of PNe , the chemically homogeneous ones and the chemically inhomogeneous ones.

For well observed chemically homogeneous PNe, Classical PNe,  it has been found that the $t^2$ values derived from $T(Balmer)$ and $T$([O III]) are similar to those derived from a) $T$(He I) and $T$([O III]), b) $T$(C II) and $T$([O III]), and  c) $T$(O II) and $T$([O III]). In addition Classical PNe have ionized masses in the $0.05$ to $0.40$  $M_{sun}$ range \citep[e.g.,][]{Mallik1988}. 

From a study of 20 well observed  Classical PNe, \citet{Peimbert2014} found the following properties: a) they do not have inclusions of high density and low temperature. b) they have $t^2$ values in the $0.024$ to $0.128$ range with a typical $t^2$ value of $0.064$, and c) they have ADF values in the $1.4$ to $10.0$ range with a typical value of 2.3. For NGC~5307 and NGC~5315 the radial velocities and the FWHM of the \ion{O}{II} and [\ion{O}{III}] emission lines are the same, indicating that they originate in the same volume elements.

PNe with inhomogeneous chemical composition have the following properties: a) they do have gaseous inclusions of high density and low temperature in the gaseous envelope, possibly the result of mass loss during close binary evolution, b) their $O/H$ values decrease from the center outward, c) they present large temperature inhomogeneities that  cannot be analyzed with the $t^2$ formalism in a one phase model, and d) their ADF values are in the $10$ to $100$ range.

\citet{Peimbert1995} found that the ratio of $C^{++}$ derived from recombination lines to that from forbidden lines amounts to about three for about 40 PNe.

Galactic chemical evolution models \citep{Carigi1995, Matteucci2006} indicate that about half of the carbon in the Galaxy has been produced by intermediate mass stars. This result is consistent with the carbon abundances derived from recombination lines, but not with the carbon abundances derived from collisionally excited lines assuming constant temperature that are typically two to three times smaller than those derived from recombination ones.

\section{\ion{H} {II} Regions}

\subsection{The $O/H$ ratio in the solar vicinity}

The brightest, closest, and most studied \ion{H}{II} region is the Orion Nebula. Under the assumption of constant temperature and based on the [\ion{O}{II}] and  [\ion{O}{III}] line intensities the $12+\log(O/H)$ value derived by different groups amounts to $8.50\pm0.02$ \citep{Peimbert1977, Osterbrock1992, Peimbert1993, Esteban1998, Deharveng2000, Esteban2004}. Alternatively \citet{Esteban2004} based on the \ion{O}{II} recombination lines obtain $12+\log(O/H)=8.72\pm0.03$. The difference between both results is real and is called the abundance discrepancy factor, ADF. We consider that the proper $O/H$ abundances are those obtained from the O and H recombination lines since their ratio is almost independent of the temperature, consequently we consider that the ADF is due to spatial temperature variations that affect the $O/H$ determinations based on O forbidden lines.

In many papers the abundances derived from forbidden lines (those of [\ion{O}{III}]) are called the direct method abundances, DM, while  those derived from recombination lines (like those of \ion{O}{II} and \ion{C}{II}) are called the temperature independent method abundances, TIM.

It is important to compare the \ion{H}{II} region abundances with the stellar abundances, since the stellar determinations are based on different methods and can help us to decide if the DM or the TIM method is the proper one to determine the $O/H$ abundances.

To compare the \ion{H}{II} region abundances with stellar abundances we have to take into account two considerations: a) we have to add to the gaseous abundances the fraction of O atoms trapped up in dust, and b) the age of the stars, because the $O/H$ abundance increases with time in the interstellar medium.

\citet{Espiritu2017} found that the fraction of oxygen atoms embedded in dust in the Orion nebula amounts to $0.126$ dex $\pm 0.024$ dex in agreement with the value of $0.12$ dex derived by \citet{Mesa-Delgado2009} and \citet{Peimbert2010}. Therefore the gaseous value derived from recombination lines plus the dust O abundance amounts to $12+\log(O/H)=8.85\pm0.05$. This result is in excellent agreement with the six most O-rich thin disk F and G dwarfs of the solar vicinity studied by \citet{Bensby2006} that show an average value of $12+\log(O/H)=8.85$. \citet{Nieva2011} have determined the $O/H$ ratio for 13 B stars of the Ori OB1 association and obtain a $12+\log(O/H)=8.77 \pm 0.03$.

A third comparison between stars and the Orion nebula can be made by using the solar $O/H$ value. The Sun has an age of 4.6 billion years and might have migrated from the place where it was born. \citet{Carigi2019} have made two chemical evolution models to fit the galactic \ion{H}{II} region abundances in our galaxy derived with the TIM and DM.
The TIM model, that fits the \ion{H}{II} region abundances for the present time,  agrees with the B stars and Cepheids abundances, while the model based on the DM abundance determinations that fits the \ion{H}{II} region abundances predicts $O/H$ values about $0.25$ dex smaller. Similarly, the $O/H$ value predicted by the TIM model for an age of $4.6$ yrs agrees with the solar abundance derived by \citet{Asplund2009}, where the probable migration of the Sun due to its age was taken into account. Alternatively the model based on the DM abundances of \ion{H}{II} regions predicts values about $0.25$ dex smaller than the solar abundances. See also Figures 7 and 9 of \citet{Carigi2019}. 

\subsection{M33 and NGC~300}

\citet{Toribio2016} from observations of \ion{H}{II} regions across the disks of NGC~300 and M33
found higher $O/H$ ratios from recombination lines, RL, than from collisionally excited lines,
CEL. For NGC 300 the RL gave abundances $0.35$ dex higher than the CEL, similarly for
M33 the RL gave abundances $0.26$ dex higher than the CEL (see their Figure 5).

\section{Primordial Helium}

From the determination of the primordial helium abundance, $Y_p$, it has
been found that temperature variations affect the determination. \citet{Peimbert2007}
find that for a given object the higher the $t^2$ value the lower the $Y_p$ that is
determined. Part of the differences between the $Y_p$ value derived by \citet{Izotov2014}
and the values derived by \citet{Peimbert2016} and \citet{Valerdi2019} are due to
the $t^2$ values adopted. The primordial helium abundance also provides strong restrictions
on the number of neutrino families and the neutron mean life.

\section{Posible causes for temperature variations}

\subsection{Deposition of mechanical energy}

\citet{Peimbert1991} suggest that the high $t^2$ values observed in PNe
and \ion{H}{II} regions might be due to the presence of shock waves. Many giant
extragalactic \ion{H}{II} regions, for example NGC~5471 \citep{Skillman1985}, and I Zw 18
\citep{Skillman1993}, present evidence of large velocity winds probably produced
by Wolf-Rayet stars and supernovae.

\citet{Gonzalez-Delgado1994} propose that stellar winds from WR stars
play an important role explaining the high $t^2$ values derived from the giant
extragalactic \ion{H}{II} region NGC~2363.

\citet{Peimbert1995} find that $T(C^{++})$, the temperature derived from the
[\ion{C} {III}] $I(1906+1909)$/\ion{C}{II} $I(4267)$, ratio is  in general considerably
smaller than the $T(O^{++})$ temperature derived from the $I(5007)/I(4363)$ ratio.
They also find that the objects with the highest $T(O^{++}) - T(C^{++})$ values are those that
show large velocities and complex velocity fields, and consequently
suggest that the deposition of mechanical energy by the stellar winds of
the PNe is the main responsible for the temperature differences.

\subsection{Chemical inhomogeneities}

\citet{Tsamis2003, Tsamis2005, Stasinska2007} suggested the presence of
metal rich droplets produced by supernova ejecta as predicted by the scenario of
\citet{Tenorio-Tagle1996}. These droplets do not get fully mixed with the
interstellar medium until they become photoionized in \ion{H}{II} regions and
they could be responsible for the ADF problem. Moreover, these droplets, if present, had to be denser and cooler than the surrounding material in H II regions.

\citet{Peimbert2013} based on high quality observations of multiplet V1 of \ion{O}{II} of 8 galactic
\ion{O}{II} regions and one extragalactic \ion{H}{II} region find that the signature of oxygen-rich
droplets of high density and low temperature is absent, ruling out the
possibility of chemical inhomogeneities in these \ion{H}{II} regions.

\citet{Gonzalez-Delgado1994} consider that some giant extragalactic \ion{H}{II} regions, like
NGC~2363, include WR stars and SNe remnants that will produce chemical inhomogeneities.

\citet{Corradi2015, GarciaRojas2016, GarciaRojas2016a, GarciaRojas2016b, Jones2016, McNabb2016, Wesson2017} and references therein have found a group of PNe that show ADF values in the $10$ to $80$ range.
This group include Haro 2-33, (Hen 2-283), Fg 1, NGC~6778, NGC~6339, Pe 1-9, MPA 1759, M1-42, Hf 2-2, Abell 30, Abell 46 and Abell 63.
These PNe correspond to binary stars that after a first ejection of gas produce a second one
with a very small mass but with a very high overabundance of O/H
relative to the first ejecta.
The ADF in these objects is strongly centrally peaked. They propose that these ADF values
should be explained in the framework of close binary evolution and
discuss the possibility that these systems could have gone through a
common envelope phase.

\subsection{Other causes of temperature inhomogeneities}

\citet{Peimbert2017} discuss other possible sources of temperature
inhomogeneities in gaseous nebulae that have been presented in the
literature: a) shadowed regions, b) spatially distributed ionization
sources, c) density inhomogeneities, d) time dependent ionization, e)
cosmic rays, f) overestimation of the intensity of weak lines, and g)
magnetic reconnection.

\section{Further Considerations}

Most \ion{H}{II} regions and PNe show temperature variations larger than
predicted  by chemically homogeneous photoionization models.

A small fraction of PNe show chemical inhomogeneities that can produce
large temperature variations.

High quality observations are needed to show if a nebula is chemically
homogeneous or not.

It is also possible to derive $t^2$ from a set of highly accurate
intensity measurements of \ion{He}{I} emission lines.

\section{Conclusions}

To study the evolution of the stars, the galaxies and the universe it is
necessary to have accurate abundances of \ion{H}{II} regions and planetary nebulae.

For chemically homogeneous nebulae the proper abundances are given by
the carbon, oxygen and hydrogen recombination lines.

The ADF values for chemically homogeneous nebulae are due to temperature
variations that can be represented by  $t^2$. The abundances derived from
collisionally excited lines, under the assumption of constant temperature
typically underestimate the abundances relative to hydrogen by a factor
of $0.25$ to $0.35$ dex.

The primordial helium abundance depends weakly on the adopted $t^2$ value,
the higher the $t^2$ value the lower the $Y_p$ value. This dependence is important for the
study of the number of neutrino families and the neutron mean life.

\acknowledgments I am grateful to Oli Dors and our Brazilian friends for organizing an
excellent meeting.

\bibliographystyle{aaabib}
\bibliography{references}

\end{document}